\documentclass[twocolumn]{autart}

\usepackage{graphics} 
\usepackage{amsmath} \usepackage{amssymb} \usepackage{amsxtra}
\usepackage{arydshln}
\usepackage{amsfonts} \usepackage[mathscr]{eucal}
\usepackage{mathrsfs} \usepackage{setspace} \usepackage{url}
\usepackage[pdftex]{graphicx} \usepackage{multirow}
\usepackage{natbib} 

\usepackage{color}

\def\Real{\mathbb{R}}

\def\Complex{\mathbb{C}}

 \def\CC{\mathcal{C}}

 \def\QQ{\mathcal{Q}}
\def\RR{\boldsymbol{\mathcal{R}}} \def\SS{\mathcal{S}}

\newcommand{\TwoOne}[2]
{\begin{bmatrix}
{#1} \\
{#2}
\end{bmatrix}
}
\newcommand{\OneTwo}[2]
{\begin{bmatrix} {#1} & {#2}
\end{bmatrix}
}
\newcommand{\TwoTwo}[4]
{\begin{bmatrix}
{#1} & {#2} \\
{#3} & {#4}
\end{bmatrix}
}

\newcommand{\bs}[1]{\boldsymbol{\mathrm{#1}}}

 \newcommand{\Hinfty}{\boldsymbol{\rm H}_{\infty}}

\begin{document}

\begin{frontmatter}

  \title{Robust stability conditions for feedback interconnections of distributed-parameter negative imaginary systems\thanksref{footnoteinfo}}

  \thanks[footnoteinfo]{This work was supported in part by the Institute for Mathematics and its Applications with funds provided by the National Science
    Foundation, the Australian Research Council, and the Swedish Research Council through the LCCC Linnaeus centre.}

  \author[kho,corr]{Sei Zhen Khong}\ead{szkhong@umn.edu}, \author[pet]{Ian R. Petersen}\ead{i.r.petersen@gmail.com}, \author[ran]{Anders Rantzer}\ead{anders.rantzer@control.lth.se}

  \thanks[corr]{Tel. +1 612 6245058. Fax +1 612 6267370.}

  \address[kho]{Institute for Mathematics and its Applications, University of Minnesota, Minneapolis, MN 55455, USA} 

  \address[pet]{ANU College of Engineering and Computer Science, Australian National University, Canberra ACT 0200, Australia}

  \address[ran]{Department of Automatic Control, Lund University, SE-221 00 Lund, Sweden}


\begin{abstract}
  Sufficient and necessary conditions for the stability of positive feedback interconnections of negative imaginary systems are derived via an
  integral quadratic constraint (IQC) approach. The IQC framework accommodates distributed-parameter systems with irrational transfer function
  representations, while generalising existing results in the literature and allowing exploitation of flexibility at zero and infinite frequencies to
  reduce conservatism in the analysis. The main results manifest the important property that the negative imaginariness of systems gives rise to a certain
  form of IQCs on positive frequencies that are bounded away from zero and infinity. Two additional sets of IQCs on the DC and instantaneous gains of
  the systems are shown to be sufficient and necessary for closed-loop stability along a homotopy of systems.
\end{abstract}

\end{frontmatter}

\section{Introduction}

The notion of negative imaginary systems was introduced in~\citep{LanPet08, PetLan10} as a natural counterpart to positive real
systems~\citep{AndVon07, Kha02, BaoLee07, Sch16}. The negative imaginary property commonly arises from the dynamics of a lightly damped structure with
colocated force actuators and position sensors (such as piezoelectric sensors)~\citep{BhiMoh09, PetLan10}. Such a system exhibits positive real
dynamics from the force input to the velocity output, but negative imaginary dynamics from the force input to the position output, whose transfer
function may be of relative degree 2. Furthermore, negative imaginary systems theory may also be employed to study certain systems that are not
passive, for which positive real results do not hold. Another area where negative imaginary dynamics can be found is that of nano-positioning
systems~\citep{DEM07}. Owing to the prevalence of negative imaginary properties in real world applications, such systems have been studied extensively
in the literature~\citep{LanPet08, PetLan10, XPL10, DPP13}. Feedback interconnections of negative imaginary systems are interpreted from a geometric
Hamiltonian systems viewpoint in~\citep{Sch11}. In \citep{WLP15}, the problem of robust output consensus of networked negative imaginary systems is
considered. Characterisations of negative imaginary systems with symmetric irrational transfer functions are considered in~\citep{FerNto13, FLN15}. A
nonlinear generalisation of negative imaginary dynamics, termed counterclockwise input-output dynamics, is given in~\citep{Ang06}.

\begin{figure}[h]
  \centering
  \includegraphics[scale=0.55]{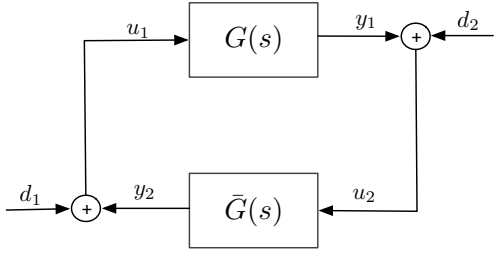}
  \caption{Positive feedback interconnection of negative imaginary systems.} \label{fig: neg_img_feedback}
\end{figure}

The robustness of feedback interconnections of open-loop stable negative imaginary systems is investigated in~\citep{LanPet08} as a parallel to the
positive real stability results~\citep{AndVon07}; see Figure~\ref{fig: neg_img_feedback}. It is shown that if the instantaneous gain of $\bar{G}$ is
positive semidefinite, i.e. $\bar{G}(\infty) \geq 0$, and the product of the instantaneous gains of $\bar{G}$ and $G$ is 0,
i.e. $G(\infty)\bar{G}(\infty) = 0$, then the closed-loop system $[G, \bar{G}]$ is internally stable if, and only if, the DC gain condition
$\bar{\lambda}(G(0)\bar{G}(0)) < 1$ is satisfied, where $\bar{\lambda}$ denotes the spectral radius. This result is further generalised
in~\citep{XPL10} to the case where $G$ may have imaginary-axis poles that are not located at the origin. Physical interpretations of these results in
terms of mass-spring-damper systems and RLC electrical networks are provided in~\citep{Pet15}. In particular, it is demonstrated using the negative
imaginary theory that certain mass-spring-damper systems with negative spring constants or RLC networks with negative inductances or capacitances are
stable, whereas the standard positive real theory is inapplicable to such non-passive systems. These stability conditions are robust in the sense that
they are invariant to negative-imaginary perturbations on the systems, provided that the aforementioned gain conditions are not violated. Stability
conditions for negative imaginary systems with poles at the origin are studied in~\citep{MKPL14}.

When the presuppositions of the stability theorems in~\citep{LanPet08, XPL10} do not hold, such as $\bar{G}(\infty)$ being sign-indefinite or
$G(\infty)\bar{G}(\infty) \neq 0$, the DC gain condition $\bar{\lambda}(G(0)\bar{G}(0)) < 1$ is not necessary. This paper derives generic sufficient
and necessary conditions for feedback stability of negative imaginary systems with respect to a specified homotopy using the theory of integral
quadratic constraints (IQC)~\citep{MegRan97, MJKR10, CJKh13}. In particular, it is established that the negative imaginary properties of the systems
give rise to complementary IQCs on a set of frequencies which do not include 0 and $\infty$ but can be arbitrarily large. This interpretation
clarifies the role of negative imaginariness in robust feedback stability analysis. Furthermore, it leads to the observation that feedback stability
follows if, and only if, there exist constant multipliers such that the corresponding complementary IQCs hold at frequencies of $0$ and $\infty$, of
which the condition in \citep{LanPet08} that $\bar{\lambda}(G(0)\bar{G}(0)) < 1$, $\bar{G}(\infty) \geq 0$, and $G(\infty)\bar{G}(\infty) = 0$ is a
special case. The robust stability result is shown to extend to negative imaginary systems that are only marginally stable, i.e. have poles on the
imaginary axis. To this end, a recently developed notion of IQCs for marginally stable systems from \citep{KLR16} is employed to conclude closed-loop
stability. This paper considers distributed-parameter linear time-invariant systems that admit irrational transfer functions. Such a class of systems
corresponds to infinite-dimensional state-space systems in the time domain~\citep{CurZwa95}. Furthermore, no explicit state-space realisations are
exploited in any of the proofs for the main results. This contrasts the preceding works~\citep{LanPet08, XPL10}, where state matrices and the negative
imaginary lemma (the counterpart to the positive real lemma) are heavily employed. Preliminary results in this direction can be found
in~\citep{KPR15}, where only sufficient IQC conditions were given for the class of proper real-rational transfer functions. Moreover, the results have
been further strengthened in this paper via the removal of an assumption on a certain residual matrix and a reconciliation with the existing results is
provided. It is noteworthy that similar necessary and sufficient IQC based results for robustness analysis involving time-delays can be found
in~\citep{Sco97} and the idea of combining IQCs which hold on subsets of the imaginary axis can be located in~\citep{JunSaf02}.

The paper evolves along the following lines. The next section introduces the notation of the paper and defines the classes of negative imaginary
systems considered. Robust stability of feedback interconnections of \emph{stable} negative imaginary systems is examined in Section~\ref{sec:
  sta}. Sufficient robust stability conditions for negative imaginary systems with imaginary-axis poles are derived in Section~\ref{sec: img}. The
necessity of IQC conditions for feedback stability of negative imaginary systems is established in Section~\ref{sec: nec_IQC}, and a reconciliation with
the existing robustness results takes place in Section~\ref{sec: relation}. Two numerical examples are given in Section~\ref{sec: ex} to
illustrate the theory. Finally, concluding remarks are provided in Section~\ref{sec: con}.

\section{Notation and preliminaries}

The notation used in this paper is defined in this section. Let $\Real$ and $\Complex$ denote, respectively, the real and complex numbers. The real
part of an $s \in \Complex$ is denoted as $\Re(s)$. $\Complex_+$ denotes the open right half plane and $\bar{\Complex}_+$ its closure.  Given an
$A \in \Complex^{m \times n}$ (resp.  $\Real^{m \times n}$), $A^* \in \Complex^{n \times m}$ (resp.  $A^T \in \Real^{n \times m}$) denotes its complex
conjugate transpose (resp. transpose). Denote by $\bar{\sigma}(A)$ and $\underline{\sigma}(A)$, the largest and smallest singular values of matrix
$A$, respectively, and by $\bar{\lambda}(B)$, the spectral radius of $B$. $I_n$ denotes the identity matrix of dimensions $n \times n$. Subsequently,
the subscript $n$ will often be dropped for simplicity.

Let $\RR^{n \times m}$ denote the set of real-rational proper transfer function matrices of dimensions $n \times m$ and
\[
\Hinfty^{n \times m} \!:=\! \left\{X : \Complex \to \Complex^{n \times m} \text{ (a.e.)} \left|
      \begin{array}{l} X \text{ is analytic
        in } \Complex_{+} \\
       \displaystyle\sup_{s \in \Complex_+} \bar{\sigma}(X(s))< \infty \end{array} \right. \!\!\right\}
\]
the set of stable transfer functions. The norm of the elements in $\Hinfty$ is denoted $\|\cdot\|_\infty$. Let $\bs{C}$ be the class of functions
$f : \Complex \to \Complex^{n \times m}$ (a.e.) that are continuous on $j\Real \cup \{\infty\}$, and $\bs{S} := \Hinfty \cap \bs{C}$. The positive
feedback interconnection of two transfer functions $G$ and $\bar{G}$, denoted by $[G, \bar{G}]$, is described by:
\[
\TwoOne{d_1}{d_2} = \TwoTwo{I}{-\bar{G}}{-G}{I}\TwoOne{u_1}{u_2};
\]
see Figure~\ref{fig: neg_img_feedback}.

\begin{defn}
  A positive feedback interconnection of $G$ and $\bar{G}$ is said to be \emph{internally stable} if
\begin{align*}
\TwoTwo{I}{-\bar{G}}{-G}{I}^{-1} = \TwoTwo{I + \bar{G}(I - G\bar{G})^{-1}G}{\bar{G}(I - G\bar{G})^{-1}}{(I - G\bar{G})^{-1}G}{(I - G\bar{G})^{-1}}
\end{align*}
is an element in $\Hinfty$.
\end{defn}

Define
\begin{align*}
  \hat{\bs{N}} & := \{R \in \bs{S}^{n \times n} : \\
               & \qquad j[R(j\omega) - R(j\omega)^*] \geq 0 \; \forall \omega \in (0, \infty)\} \quad \text{and} \\
  \bs{N}_s & := \{R \in \bs{S}^{n \times n} : \\
               & \qquad j[R(j\omega) - R(j\omega)^*] > 0 \; \forall \omega \in (0, \infty)\} \subset \hat{\bs{N}}.
\end{align*}
$\hat{\bs{N}}$ denotes the set of \emph{stable negative imaginary} transfer functions, while $\bs{N}_s$ denotes the set of \emph{strictly negative
  imaginary} transfer functions. The set of stable (strictly) negative imaginary real-rational proper transfer functions defined in~\citep{LanPet08}
is a subclass of $\hat{\bs{N}}$ ($\bs{N}_s$). In particular, an $R \in \RR \cap \hat{\bs{N}}$ satisfies $R(0) = R(0)^T \in \Real^{n \times n}$ and
$R(\infty) = R(\infty)^T \in \Real^{n \times n}$~\citep[Lem. 2]{LanPet08}. Therefore, it follows that $j[R(j\omega) - R(j\omega)^*] = 0$ when
$\omega = 0$ or $\omega = \infty$. The set of \emph{negative imaginary} transfer functions is defined below.

\begin{defn} \label{def: NI}
A transfer function $R : \Complex \to \Complex^{n \times n}$ (a.e.) is said to be \emph{negative imaginary} if
\begin{enumerate} \renewcommand{\theenumi}{\textup{(\roman{enumi})}}\renewcommand{\labelenumi}{\theenumi}
 \item $R$ is analytic in $\Re(s) > 0$ and has no singularities at $s = 0$ and $s = \infty$;

\item $R(\cdot + \epsilon) \in \Hinfty$ for some $\epsilon > 0$;

 \item $R$ has at most a finite number of singularities on the imaginary axis and they appear as complex conjugate pairs;

 \item $R(j\omega)$ is continuous on $j\Real$ except when $j\omega$ is a singularity;

 \item $j[R(j\omega) - R(j\omega)^*] \geq 0$ for all $\omega \in (0, \infty)$ except values $\omega$ where $j\omega$ is a singularity of $R(s)$;

 \item if $s = j\omega_0$ with $\omega_0 \in (0, \infty)$ is a singularity of $R(s)$, then it is a simple pole and the residue matrix
   $\displaystyle\lim_{s \rightarrow j\omega_0} (s - j\omega_0) jR(s)$ is Hermitian and positive semidefinite.
\end{enumerate}
\end{defn}

Denote by $\bs{N}$ the set of negative imaginary transfer functions. Notice that $\bs{N}_s \subset \hat{\bs{N}} \subset \bs{N}$. The set of negative
imaginary real-rational proper transfer functions defined in~\citep{XPL10} is a subclass of $\bs{N}$. It is noted here that the definitions given above
differ from those in~\citep{FerNto13, FLN15} in that no efforts have been made here to link the definitions to the classical positive real
theory. Instead, only the properties crucial to the robust stability conditions to be derived in the next sections are stipulated.

In order to accommodate possibly irrational transfer functions with poles on the imaginary axis in this paper, some background material on IQCs needs
to be stated. To this end, the following notational definitions are important. Given an $\epsilon > 0$ and a point $jq \in j\Real$, define the
semi-circle of radius $\epsilon$ in the right-half plane as
\[
\SS_\epsilon(jq) := \{s \in \Complex : |s - jq| = \epsilon, \Re(s) > 0\}
\]
and $\SS_0(jq) := \{\}$. Given a finite ordered set $j\QQ = \{jq_1, jq_2, \ldots, jq_K\}\subset j\Real$ with $q_1 > q_2 > \ldots > q_K$, define a contour parameterised by $\epsilon \geq 0$ as
\begin{align} \label{eq: ind_contour}
\begin{split}
\CC_\epsilon(j\QQ) := j[q_1 + \epsilon, \infty) & \cup \SS_\epsilon(jq_1) \cup j[q_2 + \epsilon, q_1 - \epsilon] \\
&  \cup \SS_\epsilon(jq_2)   \cup j[q_3 + \epsilon, q_2 - \epsilon] \\
& \qquad\qquad \vdots \\
&  \cup \SS_\epsilon(jq_K) \cup  j(-\infty, q_K - \epsilon].
\end{split}
\end{align}
that is, a straight line on the imaginary axis indented to the right of every point in $j\QQ$ by a semi-circle of radius $\epsilon$. In particular, notice that $C_0(j\QQ) = j\Real$ for any $j\QQ
\subset j\Real$. Denote by $\CC_\epsilon^+(j\QQ)$ the open half plane that lies to the right of $\CC_\epsilon(j\QQ)$ defined in (\ref{eq: ind_contour}), i.e.
\[
\CC_\epsilon^+(j\QQ) := \{s = \sigma + j\omega \in \Complex\ |\ \bar{\sigma} + j\omega \in \CC_\epsilon(j\QQ) \Rightarrow \sigma > \bar{\sigma} \},
\]
and $\bar{\CC}_\epsilon^+(j\QQ)$ its closure. Let $\bs{C}_\epsilon(j\QQ)$ be the class of functions continuous on
$\CC_\epsilon(j\QQ) \cup \{\infty\}$. Given $X \in \bs{C}_\epsilon(j\QQ)^{n \times m}$, define
$\|X\|_{\bs{C}_\epsilon(j\QQ)} := \sup_{s \in \CC_\epsilon(j\QQ)} \bar{\sigma}(X(s))$. Let $\bs{S}_\epsilon(j\QQ)$ be the subclass of
$\bs{C}_\epsilon(j\QQ)$ containing functions that have analytic continuation into $\CC_\epsilon^+(j\QQ)$. Note that
$\bs{S} = \bs{S}_0(j\QQ) \subset \bs{S}_\epsilon(j\QQ)$ for all $\epsilon \geq 0$.

The following result can be established using the arguments in~\citep[Thm. 4.4 and 4.5]{KLR16}. It has further been generalised in~\citep{KLK17}.

\begin{prop} \label{prop: IQC} Given $G \in \bs{S}_\epsilon(j\QQ)$ and $\bar{G} \in \bs{S}_\epsilon(j\QQ)$, the closed-loop transfer
  function of the feedback interconnection $[G, \bar{G}]$
  \[
  H := \TwoTwo{I}{-\bar{G}}{-G}{I}^{-1} \in \bs{S}_\epsilon(j\QQ)
  \]
  for all $\epsilon > 0$, if there exist a bounded $\Pi \in \bs{C}^{(n + m)\times (n + m)}$ and $\eta > 0$ such that $\Pi(j\omega) = \Pi(j\omega)^*$
  for all $\omega \in \Real$ and the following complementary IQC conditions hold:
  \begin{enumerate} \renewcommand{\theenumi}{\textup{(\roman{enumi})}}\renewcommand{\labelenumi}{\theenumi}
  \item for all $\omega \in \Real \setminus \QQ$,
    \[
    \TwoOne{\bar{G}(j\omega)}{I}^* \Pi(j\omega) \TwoOne{\bar{G}(j\omega)}{I} \leq -\eta I;
    \]

  \item for all $\omega \in \Real \setminus \QQ, \tau \in [0, 1]$,
    \[
    \TwoOne{I}{\tau G(j\omega)}^* \Pi(j\omega) \TwoOne{I}{\tau G(j\omega)} \geq 0.
    \]
  \end{enumerate}
Furthermore, when $j\QQ = \{\}$, $\epsilon$ can be taken to be $0$, whereby $\bs{S}_0(\{\}) = \bs{S} \subset \Hinfty$.
\end{prop}

\section{Stable negative imaginary transfer functions} \label{sec: sta}

In this section, sufficient IQC conditions which guarantee closed-loop stability with stable negative imaginary systems are derived. The proof methods
in this section will be reused in the subsequent section where systems with nonzero imaginary-axis poles are accommodated. They involve constructing a
3-part IQC, each of which is valid for a different frequency range. An intuitive depiction of the theorem below is provided after the proof.

\begin{thm} \label{thm: stable} Given $G \in \hat{\bs{N}}^{n \times n}$ and $\bar{G} \in \bs{N}_s^{n \times n}$, suppose there exist
  $\Pi_0 = \Pi_0^* \in \Complex^{2n \times 2n}$, $\Pi_\infty = \Pi_\infty^* \in \Complex^{2n \times 2n}$ such that for some $\bar{\epsilon} > 0$ and
  all $\tau \in [0, 1]$,
  \begin{align} \label{eq: zero_freq}
    \begin{split}
      \TwoOne{\bar{G}(0)}{I}^* \Pi_0 \TwoOne{\bar{G}(0)}{I} & \leq -\bar{\epsilon} I; \\
      \TwoOne{I}{\tau G(0)}^* \Pi_0 \TwoOne{I}{\tau G(0)} & \geq 0
    \end{split}
  \end{align}
  and
  \begin{align} \label{eq: inf_freq}
    \begin{split}
      \TwoOne{\bar{G}(\infty)}{I}^* \Pi_\infty \TwoOne{\bar{G}(\infty)}{I} & \leq -\bar{\epsilon} I ; \\
      \TwoOne{I}{\tau G(\infty)}^* \Pi_\infty \TwoOne{I}{\tau G(\infty)} & \geq 0.
    \end{split}
  \end{align}
  Then the feedback interconnection $[\tau G, \bar{G}]$ is internally stable for all $\tau \in [0, 1]$.
\end{thm}

\begin{pf}
  Let $\hat{\Pi}_0 := 2\Pi_0 + \mu I$ and $\hat{\Pi}_\infty := 2\Pi_\infty + \mu I$. Then for sufficiently small $\mu > 0$, the inequalities
  \eqref{eq: zero_freq} and \eqref{eq: inf_freq} imply, respectively,
  \begin{align*}
    \TwoOne{\bar{G}(0)}{I}^* \hat{\Pi}_0 \TwoOne{\bar{G}(0)}{I} & \leq -\epsilon I; \\
    \TwoOne{I}{\tau G(0)}^* \hat{\Pi}_0 \TwoOne{I}{\tau G(0)} & \geq \epsilon I
  \end{align*}
  and
  \begin{align*}
    \TwoOne{\bar{G}(\infty)}{I}^* \hat{\Pi}_\infty \TwoOne{\bar{G}(\infty)}{I} & \leq -\epsilon I ; \\
    \TwoOne{I}{\tau G(\infty)}^* \hat{\Pi}_\infty \TwoOne{I}{\tau G(\infty)} & \geq \epsilon I
  \end{align*}
  for some $\epsilon > 0$ and all $\tau \in [0, 1]$.  By the continuity of $G \in \bs{S}$ and $\bar{G} \in \bs{S}$ on $j\Real$, these
  imply there exist sufficiently small $\underline{\Omega} > 0$ and sufficiently large $\bar{\Omega} > 0$ such that
  \begin{align} \label{eq: small_freq}
    \begin{split}
      \TwoOne{\bar{G}(j\omega)}{I}^* \hat{\Pi}_0 \TwoOne{\bar{G}(j\omega)}{I} & \leq -\frac{\epsilon}{2} I; \\
      \TwoOne{I}{\tau G(j\omega)}^* \hat{\Pi}_0 \TwoOne{I}{\tau G(j\omega)} & \geq \frac{\epsilon}{2} I
    \end{split}
  \end{align}
  for all $\omega \in [0, \underline{\Omega}]$, $\tau \in [0, 1]$ and
  \begin{align} \label{eq: large_freq}
    \begin{split}
      \TwoOne{\bar{G}(j\omega)}{I}^* \hat{\Pi}_\infty \TwoOne{\bar{G}(j\omega)}{I} & \leq -\frac{\epsilon}{2} I \\
      \TwoOne{I}{\tau G(j\omega)}^* \hat{\Pi}_\infty \TwoOne{I}{\tau G(j\omega)} & \geq \frac{\epsilon}{2} I
    \end{split}
  \end{align}
  for all $\omega \in [\bar{\Omega}, \infty]$, $\tau \in [0, 1]$. Now note from the definitions of $\hat{\bs{N}}$ and $\bs{N}_s$ that $G \in \hat{\bs{N}}$
  and $\bar{G} \in \bs{N}_s$ implies that 
  \begin{align} \label{eq: mid_freq2}
\begin{split}
    \TwoOne{\bar{G}(j\omega)}{I}^* \Pi_m \TwoOne{\bar{G}(j\omega)}{I} & \leq 0 ; \\
    \TwoOne{I}{\tau G(j\omega)}^* \Pi_m \TwoOne{I}{\tau G(j\omega)} & \geq 0
\end{split}
  \end{align}
  for all $\tau \in [0, 1]$ and $\omega \in [0, \infty)$, where
  \[
  \Pi_m := \TwoTwo{0}{j I}{-j I}{0}.
  \] 
  Furthermore, there exists $\bar{\eta} > 0$ such that
  \begin{align} \label{eq: mid_freq1}
\begin{split}
    \TwoOne{\bar{G}(j\omega)}{I}^* \Pi_m \TwoOne{\bar{G}(j\omega)}{I} & \leq -\bar{\eta} I ; \\
    \TwoOne{I}{\tau G(j\omega)}^* \Pi_m \TwoOne{I}{\tau G(j\omega)} & \geq 0
\end{split}
  \end{align}
  for all $\tau \in [0, 1]$ and $\omega \in [\underline{\Omega}, \bar{\Omega}]$.  Let $\hat{\Pi}_m := 2\Pi_m + \bar{\mu} I$, then for sufficiently
  small $\bar{\mu} > 0$, \eqref{eq: mid_freq1} implies that
  \begin{align} \label{eq: mid_freq}
    \begin{split}
      \TwoOne{\bar{G}(j\omega)}{I}^* \hat{\Pi}_m \TwoOne{\bar{G}(j\omega)}{I} & \leq -\eta I ; \\
      \TwoOne{I}{\tau G(j\omega)}^* \hat{\Pi}_m \TwoOne{I}{\tau G(j\omega)} & \geq \eta I
    \end{split}
  \end{align}
  for some $\eta > 0$, all $\tau \in [0, 1]$ and $\omega \in [\underline{\Omega}, \bar{\Omega}]$.  Define
  \[
  \gamma_0(j\omega) := \left\{\begin{array}{ll}
                                1 & \quad \omega \in [-\underline{\Omega}, \underline{\Omega}] \\
                                0 & \quad \text{otherwise},
                              \end{array}\right.
                            \]
                            \[
                            \gamma_\infty(j\omega) := \left\{\begin{array}{ll}
                                                               1 & \quad \omega \in [\bar{\Omega}, \infty) \cup (-\infty, -\bar{\Omega}]\\
                                                               0 & \quad \text{otherwise},
                                                             \end{array}\right.
                                                           \]
                                                           and
                                                           \begin{align} \label{eq: fab_pi} \Pi(j\omega) := \gamma_0(j\omega)\hat{\Pi}_0 + \hat{\Pi}_m
                                                             + \gamma_\infty(j\omega)\hat{\Pi}_\infty, \omega \in \Real.
                                                           \end{align}
                                                           Combining \eqref{eq: small_freq}, \eqref{eq: large_freq}, \eqref{eq: mid_freq2}, and
                                                           \eqref{eq: mid_freq} yields that
                                                           \begin{align} \label{eq: fab_IQC}
                                                             \begin{split} 
                                                               \TwoOne{\bar{G}(j\omega)}{I}^* \Pi(j\omega) \TwoOne{\bar{G}(j\omega)}{I} & \leq -\zeta I ; \\
                                                               \TwoOne{I}{\tau G(j\omega)}^* \Pi(j\omega) \TwoOne{I}{\tau G(j\omega)} & \geq 0
                                                             \end{split}
                                                           \end{align}
                                                           for all $\omega \in [0, \infty)$, $\tau \in [0, 1]$ and some $\zeta > 0$. By taking the
                                                           complex conjugate on both sides of the inequalities, observe that \eqref{eq: fab_IQC} holds
                                                           for all $\omega \in \Real$. The stability of $[\tau G, \bar{G}]$ for $\tau \in [0, 1]$ then
                                                           follows from as a special case of the IQC result in Proposition~\ref{prop: IQC} with
                                                           $j\QQ = \{\}$ and $\epsilon = 0$. \hfill $\qed$
                                                         \end{pf}

                                                         The stability conditions in Theorem~\ref{thm: stable} are robust in the sense that if
                                                         $[G, \bar{G}]$ is stable, then the feedback interconnection remains stable with respect to
                                                         negative imaginary perturbations on $G$ and $\bar{G}$ that do not violate the DC and
                                                         instantaneous gain conditions~\eqref{eq: zero_freq} and~\eqref{eq: inf_freq}.

 \begin{rem}
   The proof of Theorem~\ref{thm: stable} relies on the homotopy approach and the fact that $[\tau G, \bar{G}]$ is stable when $\tau = 0$. In
   particular, perturbation arguments can be used to show that $[\tau G, \bar{G}]$ is stable as one increases $\tau$ from $0$ to $1$. In the case
   where $\|G^{-1}\|_\infty^{-1}$ and $\|\bar{G}^{-1}\|_\infty^{-1}$ are both finite, it follows from the large-gain theorem in~\cite{ZMC08} that
   $[\tau G, \bar{G}]$ is stable for sufficiently large $\tau$. This allows the use of an alternative homotopy, whereby Theorem~\ref{thm: stable} can
   be modified with $\tau \in [0, 1]$ replaced by $\tau \geq 1$.
\end{rem}

                                                         \begin{rem} \label{rem: mod_IQC} Note from its proof that the conclusion in Theorem~\ref{thm:
                                                             stable} remains valid with \eqref{eq: zero_freq} and \eqref{eq: inf_freq} modified into
                                                           the following forms:
                                                           \begin{align}
                                                             \begin{split} \label{eq: mod_IQC0}
                                                               \TwoOne{\tau \bar{G}(0)}{I}^* \Pi_0 \TwoOne{\tau \bar{G}(0)}{I} & \geq \epsilon I; \\
                                                               \TwoOne{I}{G(0)}^* \Pi_0 \TwoOne{I}{G(0)} & \leq 0
                                                             \end{split}
                                                           \end{align}
                                                           and
                                                           \begin{align} \label{eq: mod_IQC_inf}
                                                             \begin{split}
                                                               \TwoOne{\tau \bar{G}(\infty)}{I}^* \Pi_\infty \TwoOne{\tau \bar{G}(\infty)}{I} & \geq \epsilon I ; \\
                                                               \TwoOne{I}{G(\infty)}^* \Pi_\infty \TwoOne{I}{G(\infty)} & \leq 0.
                                                             \end{split}
                                                           \end{align}
                                                         \end{rem}

  Theorem~\ref{thm: stable} is established by fabricating a 3-part multiplier $\Pi$ in \eqref{eq: fab_pi} in such a way that the standard IQC result can be
  applied to conclude closed-loop stability. In particular, the fact that $G \in \hat{\bs{N}}$ and $\bar{G} \in \bs{N}_s$ implies the complementary
  IQC inequalities hold for positive frequencies that are bounded from zero and infinity; see \eqref{eq: mid_freq}. The additional matrix inequalities
  \eqref{eq: zero_freq} and \eqref{eq: inf_freq} in the theorem imply the complementary IQC inequalities for sufficiently small and sufficiently large
  frequencies, i.e. \eqref{eq: small_freq} and \eqref{eq: large_freq} respectively.

\begin{cor} \label{cor: sta} Given $G \in \hat{\bs{N}}$ and $\bar{G} \in \bs{N}_s$, suppose $\bar{\sigma}(G(0)\bar{G}(0)) < 1$ and
  $\bar{\sigma}(G(\infty)\bar{G}(\infty)) < 1$, then the feedback interconnection $[\tau G, \bar{G}]$ is internally stable for all $\tau \in [0, 1]$.
\end{cor}

\begin{pf}
  Note that the hypothesis is equivalent to $\bar{\lambda}(\bar{G}(j\omega)^*G(j\omega)^*G(j\omega)\bar{G}(j\omega)) < 1$ for $\omega = 0$ and
  $\omega = \infty$. It follows that the matrix inequalities \eqref{eq: zero_freq} and \eqref{eq: inf_freq} in Theorem~\ref{thm: stable} hold with
  respect to
  \[
  \Pi_0 := \TwoTwo{G(0)^*G(0)}{0}{0}{-I}
  \]
  and
  \[
  \Pi_\infty := \TwoTwo{G(\infty)^*G(\infty)}{0}{0}{-I},
  \]
  as required. \hfill$\qed$
\end{pf}

\section{Negative imaginary transfer functions with imaginary-axis poles} \label{sec: img}

IQC-based conditions for feedback stability of negative imaginary systems with imaginary-axis poles are established in this section. The proofs rely
on the arguments detailed in Section~\ref{sec: sta}.  In order to accommodate negative imaginary transfer functions with imaginary-axis poles, the
generalised version of the IQC result, i.e. Proposition~\ref{prop: IQC}, is needed. 

\begin{thm} \label{thm: FB_ori} Given $\bar{G} \in \bs{N}_s$ and $G \in \bs{N}$, suppose there exist $\Pi_0$ and $\Pi_\infty$ such that for all
  $\tau \in [0, 1]$,
\begin{align} \label{eq: zero_f}
\begin{split} 
\TwoOne{\bar{G}(0)}{I}^* \Pi_0 \TwoOne{\bar{G}(0)}{I} & < 0; \\
\TwoOne{I}{\tau G(0)}^* \Pi_0 \TwoOne{I}{\tau G(0)} & \geq 0,
\end{split}
\end{align}
\begin{align} \label{eq: inf_f}
\begin{split}
\TwoOne{\bar{G}(\infty)}{I}^* \Pi_\infty \TwoOne{\bar{G}(\infty)}{I} & < 0; \\
\TwoOne{I}{\tau G(\infty)}^* \Pi_\infty \TwoOne{I}{\tau G(\infty)} & \geq 0,
\end{split}
\end{align}
  then the feedback interconnection $[\tau G, \bar{G}]$ is internally stable for all $\tau \in [0, 1]$.
\end{thm}

\begin{pf}
  The same arguments in the proof for Theorem~\ref{thm: stable} can be used to establish \eqref{eq: fab_IQC} for all $\tau \in [0, 1]$ and $\omega \in [0, \infty] \setminus \QQ$, where $j\QQ$ denotes
  the set of imaginary-axis poles of $G$. The only additional requirement is that $\underline{\Omega}$ needs to be sufficiently small and $\bar{\Omega}$ sufficiently large so that $\QQ \cap [0,
  \underline{\Omega}] = \{\}$ and $\QQ \cap [\bar{\Omega}, \infty] = \{\}$. By Proposition~\ref{prop: IQC}, this then implies that the closed-loop transfer function
\begin{align} \label{eq: CL_TF}
H_\tau := \TwoTwo{I}{-\bar{G}}{-\tau G}{I}^{-1} \in \bs{S}_\epsilon(j\QQ)
\end{align}
for all $\tau \in [0, 1]$ and $\epsilon > 0$. In what follows, we show that $H_\tau$ has also no poles in $j\QQ$, which then implies
$H_\tau \in \bs{S} \subset \Hinfty$, i.e. the feedback interconnection $[\tau G, \bar{G}]$ is stable, for all $\tau \in [0, 1]$.

First note that $\bar{G} \in \bs{N}_s$ implies $\det(\bar{G}(j\omega)) \neq 0$ for $\omega \neq 0$. To see this, observe that if $\det(\bar{G}(j\omega_0)) = 0$ for some $\omega_0 > 0$, then there exists $v \in
\Complex^n$ such that $\bar{G}(j\omega_0)v = 0$. It follows that $v^* j[\bar{G}(j\omega_0) - \bar{G}(j\omega_0)^*] v = 0$, which violates the supposition that $\bar{G} \in \bs{N}_s$. As such, 
\begin{align} \label{eq: detM}
\det(\bar{G}(j\omega)) \neq 0 \quad \forall \omega \in (0, \infty). 
\end{align}
This implies that there is no closed right-half plane pole-zero cancellation in the product $\tau G(s)\bar{G}(s)$, since $\tau G \in \bs{N}$ has no poles at the
origin. Therefore, for any $\omega_0 \in \Real$, $j\omega_0$ is a pole of $H_\tau(s)$ if, and only if, $j\omega_0$ is a pole of
$(I - \tau G(s)\bar{G}(s))^{-1}$.

Now let $j\omega_0$, $\omega_0 > 0$ be an imaginary-axis pole of $G \in \bs{N}$, i.e. $j\omega_0 \in j\QQ$. Suppose to the contrapositive that
$j\omega_0$ is a pole of $(I - \tau G(s)\bar{G}(s))^{-1}$ for some $\tau \in [0, 1]$. Then it follows that $\det(I - \tau
G(j\omega)\bar{G}(j\omega))$, or
\[
\det(-\bar{G}(j\omega)^{-1} + \tau G(j\omega) + \bar{G}(j\omega)^{-*} - \tau G(j\omega)^*),
\]
can be made arbitrarily small by having $|\omega - \omega_0|$ sufficiently small. However, by the negative imaginary properties of $G$ and $\bar{G}$, it holds that
\[
j[\tau G(j\omega) - \tau G(j\omega)^*] \geq 0
\]
and
\[
j[\bar{G}(j\omega) - \bar{G}(j\omega)^*] \geq \eta > 0,
\]
or equivalently,
\[
j[\bar{G}(j\omega)^{-*} - \bar{G}(j\omega)^{-1}] \geq \bar{\eta} > 0,
\]
leading to a contradiction. Therefore, $S(s)^{-1}$ has no pole at $j\omega_0$. This implies that $(I - \tau G(s)\bar{G}(s))^{-1}$, and hence
$H_\tau(s)$, has no pole at every $j\omega_0 \in j\QQ$.

All in all, the closed-loop transfer function $H_\tau(s) \in \bs{S} \subset \Hinfty$, and the feedback interconnection $[\tau G, \bar{G}]$ is stable for all
$\tau \in [0, 1]$. \hfill $\qed$
\end{pf}

\begin{rem} \label{rem: mod_IQC2} Note from its proof that the conclusion of Theorem~\ref{thm: FB_ori} remains valid with \eqref{eq: zero_f} and
  \eqref{eq: inf_f} written as \eqref{eq: mod_IQC0} and \eqref{eq: mod_IQC_inf}, respectively.
\end{rem}

\begin{cor} \label{cor: img} Given $\bar{G} \in \bs{N}_s$ and $G \in \bs{N}$, suppose $\bar{\sigma}(G(0)\bar{G}(0)) < 1$,
  $\bar{\sigma}(G(\infty)\bar{G}(\infty)) < 1$, and for every $j\omega_0$, $\omega_0 > 0$, that is a pole of $G$, the residue matrix
  $\displaystyle\lim_{s \rightarrow j\omega_0} (s - j\omega_0) jG(s)$ is positive definite.  Under these conditions, the feedback interconnection
  $[\tau G, \bar{G}]$ is internally stable for all $\tau [0, 1]$.
\end{cor}

\begin{pf}
  The same arguments in the proof for Corollary~\ref{cor: sta} can be applied to show that the conditions in Theorem~\ref{thm: FB_ori} hold, which
  implies the claim. \hfill $\qed$
\end{pf}

\section{Necessity of IQCs for closed-loop stability} \label{sec: nec_IQC}

In this section, necessity of the IQC conditions for robust stability of positive feedback interconnections of negative imaginary systems is
established. First, the following lemma, which can be found in~\cite[Cor. 1]{IwaHar98}, is stated. An alternative proof is provided below for completeness.

\begin{lem} \label{lem: coprime_IQC}
 Given $A, B \in \Complex^{n \times n}$, if $I - \tau BA$ is nonsingular for all $\tau \in [0, 1]$, then there exists a $\Pi = \Pi^* \in \Complex^{2n
   \times 2n}$ such that
  \begin{align}
      \TwoOne{\tau A}{I}^* \Pi \TwoOne{\tau A}{I} & > 0 \quad \forall \tau \in [0, 1]; \label{eq: A} \\
      \TwoOne{I}{B}^* \Pi \TwoOne{I}{B} & \leq 0. \label{eq: B}
  \end{align}
In particular, such a $\Pi$ may be taken as
\begin{align*}
\Pi & := \tilde{Y}^*\tilde{Y} \\
\tilde{Y} & := \OneTwo{-(I + B^*B)^{-\frac{1}{2}}B}{(I + B^*B)^{-\frac{1}{2}}}.
\end{align*}
\end{lem}

\begin{pf}
Define 
\[
Y := \TwoOne{(I + B^*B)^{-\frac{1}{2}}}{B(I + B^*B)^{-\frac{1}{2}}}
\]
and note that $Y^* \Pi Y = 0$. Multiplying by $(I + B^*B)^{\frac{1}{2}}$ from the left and right, this implies \eqref{eq: B}.

Define for $\tau \in [0, 1]$
\begin{align*}
\tilde{X}_\tau & := \OneTwo{(I +\tau^2 A^*A)^{-\frac{1}{2}}}{-(I + \tau^2 A^*A)^{-\frac{1}{2}}\tau A} \\
X_\tau & := \TwoOne{\tau A(I + \tau^2 A^*A)^{-\frac{1}{2}}}{(I + \tau^2 A^*A)^{-\frac{1}{2}}}.
\end{align*}
By the hypothesis that $I - \tau BA$ is nonsingular for all $\tau \in [0, 1]$, we have
\begin{align*}
0 & < 1\left/\bar{\sigma}\left(\TwoOne{\tau A}{I} (I - \tau BA)^{-1} \OneTwo{-B}{I} \right)\right. \\
& = 1\left/\bar{\sigma}  \left(\TwoOne{\tau A(I + \tau^2 A^*A)^{-\frac{1}{2}}}{(I + \tau^2 A^*A)^{-\frac{1}{2}}} \right.\right. \times \\
& \qquad \left((I + B^*B)^{-\frac{1}{2}}(I + \tau^2 A^*A)^{-\frac{1}{2}}\right. \\
& \qquad\quad \left. - (I + B^*B)^{-\frac{1}{2}}\tau BA (I + \tau^2 A^*A)^{-\frac{1}{2}}\right)^{-1} \times  \\
& \qquad \left. \OneTwo{-(I + B^*B)^{-\frac{1}{2}}B}{(I + B^*B)^{-\frac{1}{2}}}\right) \\
& = 1\left/\bar{\sigma}\left(X_\tau (\tilde{Y} X_\tau)^{-1} \tilde{Y} \right)\right. \\
& = 1\left/\bar{\sigma}\left((\tilde{Y} X_\tau)^{-1} \right)\right. \\
& = \underline{\sigma}\left((\tilde{Y} X_\tau)\right),
\end{align*}
where the fact that $X_\tau^*X_\tau = I$ and $\tilde{Y}\tilde{Y}^* = I$ has been used in the second equality. Now note that
$\underline{\sigma}\left((\tilde{Y} X_\tau)\right) > 0 \; \forall \tau [0, 1]$ implies
\[
X_\tau^* \tilde{Y}^*\tilde{Y} X_\tau = X_\tau^* \Pi X_\tau > 0.
\]
Multiplying from the left and right by $(I + \tau^2 A^*A)^{\frac{1}{2}}$ gives \eqref{eq: A}, as required. \hfill $\qed$
\end{pf}

The following result on the sufficiency and necessity of IQCs is in order.

\begin{thm} \label{thm: nec_IQC} Given $\bar{G} \in \bs{N}_s$ and $G \in \bs{N}$, the feedback interconnection $[\tau G, \bar{G}]$ is internally
  stable for all $\tau \in [0, 1]$ if, and only if, there exist $\Pi_0 = \Pi_0^* \in \Complex^{2n \times 2n}$ and
  $\Pi_\infty = \Pi_\infty^* \in \Complex^{2n \times 2n}$ such that
  \begin{align*}
    \begin{split}
      \TwoOne{\tau \bar{G}(0)}{I}^* \Pi_0 \TwoOne{\tau \bar{G}(0)}{I} & > 0; \\
      \TwoOne{I}{G(0)}^* \Pi_0 \TwoOne{I}{G(0)} & \leq 0
    \end{split}
  \end{align*}
  and
  \begin{align*}
    \begin{split}
      \TwoOne{\tau \bar{G}(\infty)}{I}^* \Pi_\infty \TwoOne{\tau \bar{G}(\infty)}{I} & > 0; \\
      \TwoOne{I}{G(\infty)}^* \Pi_\infty \TwoOne{I}{G(\infty)} & \leq 0
    \end{split}
  \end{align*}
for all $\tau \in [0, 1]$.
\end{thm}

\begin{pf}
  Sufficiency has been shown in Theorem~\ref{thm: FB_ori}. To establish necessity, note that the stability of $[\tau G, \bar{G}]$ for all
  $\tau \in [0, 1]$ implies that $(I - \tau G\bar{G})^{-1} \in \Hinfty$, which in turn implies that $I - \tau G(\omega)\bar{G}(\omega)$ is
  nonsingular for $\omega = 0$ and $\omega = \infty$. The claim can then established by invoking Lemma~\ref{lem: coprime_IQC} for these two
  frequencies. \hfill $\qed$
\end{pf}


\section{Reconciliation with existing results} \label{sec: relation}

In this section, we generalise the sufficiency part of the main result in~\citep{LanPet08} to irrational transfer functions via the IQC approach
described in the preceding sections. In doing so, it can be observed that the sufficiency part of~\citep[Thm. 5]{LanPet08} is a special case of that
of Theorem~\ref{thm: nec_IQC}.


\begin{prop}
  Given $\bar{G} \in \bs{N}_s$ and $G \in \hat{\bs{N}}$, then $[\tau G, \bar{G}]$ is internally stable for all $\tau \in [0, 1]$ if
  $\bar{G}(\infty) \geq 0$ and $G(\infty)\bar{G}(\infty) = 0$, and $\bar{\lambda}(G(0)\bar{G}(0)) < 1$.
\end{prop}

\begin{pf}
 First note that $\bar{\lambda}(G(0)\bar{G}(0)) < 1$ implies 
\[
\bar{\lambda}(\tau G(0)\bar{G}(0)) < 1 \quad \text{for all } \tau \in [0, 1]. 
\]
This in turn implies that $(I - \tau(G(0)\bar{G}(0))$ is nonsingular for all $\tau \in [0, 1]$. To see this, suppose to the contrapositive that
$(I - \tau(G(0)\bar{G}(0))$ is singular. This implies that there exists a $v \in \Complex^n$ such that $(I - \tau(G(0)\bar{G}(0))v = 0$, or
$\tau(G(0)\bar{G}(0)v = v$, i.e. $v$ is an eigenvector of $\tau(G(0)\bar{G}(0)$ corresponding to the eigenvalue $1$. This contradicts the fact that
$\bar{\lambda}(\tau G(0)\bar{G}(0)) < 1$. Thus, by invoking Lemma~\ref{lem: coprime_IQC}, there exists $\Pi_0 = \Pi_0^*$ such that \eqref{eq:
  mod_IQC0} in Remarks~\ref{rem: mod_IQC2} and \ref{rem: mod_IQC} holds.

By defining
\[
\Pi_\infty := \TwoTwo{-G(\infty)^*G(\infty)}{0}{0}{I},
\]
it follows that 
\[
\TwoOne{\tau \bar{G}(\infty)}{I}^* \Pi_\infty \TwoOne{\tau \bar{G}(\infty)}{I} = I > 0
\]
for all $\tau \in [0, 1]$ and
\begin{align*}
\TwoOne{I}{G(\infty)}^* \Pi_\infty \TwoOne{I}{G(\infty)} & = G(\infty)^*G(\infty) - G(\infty)^*G(\infty)\\
&  = 0.
\end{align*}
In other words, \eqref{eq: mod_IQC_inf} in Remarks~\ref{rem: mod_IQC2} and \ref{rem: mod_IQC} holds. Therefore, internal stability of $[G, \bar{G}]$
follows by Theorem~\ref{thm: FB_ori} and Remark~\ref{rem: mod_IQC2}. \hfill $\qed$
\end{pf}

In~\citep[Thm. 5]{LanPet08}, it is shown that if the presuppositions $\bar{G}(\infty) \geq 0$ and $G(\infty)\bar{G}(\infty) = 0$ hold, then the
internal stability of $[G, \bar{G}]$ implies $\bar{\lambda}(G(0)\bar{G}(0)) < 1$. When the former are not known in advance, Theorem~\ref{thm: nec_IQC}
presents a generalised form of necessity of the robust stability conditions in terms of IQCs.

\section{Numerical examples} \label{sec: ex}

\subsection{Rational transfer functions}

\begin{figure}[h]
  \centering
  \includegraphics[scale=0.6]{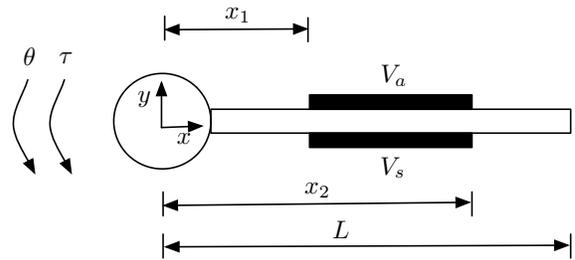}
  \caption{Schematic diagram of a slewing beam equivalent to a robotic arm.} \label{fig: beam}
\end{figure}

Consider a robotic arm pinned to a motor at one end and an equivalent slewing beam model shown in Figure~\ref{fig: beam}; see~\citep{PotAlb95}. Two
piezoelectric patches are attached to the arm on either side of the beam. They act as an actuator and a sensor, respectively. The system has input
voltage $V_a$ applied to the piezoelectric actuator and input torque $\tau$ applied by the motor. On the other hand, the outputs of the system are the
voltage $V_s$ produced by the piezoelectric sensor and the motor hub angle $\theta$. A distributed-parameter transfer function matrix for the robotic
arm is provided in~\citep{PotAlb95}:
\[
\TwoTwo{\frac{N_{\tau, \theta}(s)}{D(s)}}{\frac{N_{V_a, \theta}(s)}{D(s)}}{\frac{N_{\tau, V_s}(s)}{D(s)}}{\frac{N_{V_a, V_s}(s)}{D(s)}},
\]
where $N_{\tau, \theta}$, $N_{V_a, \theta}$, $N_{\tau, V_s}$, $N_{V_a, V_s}$, and $D$ are given in equations (26)-(28) in~\citep{PotAlb95}.

By approximating the distributed-parameter model as in~\citep{MKPL14} with a first-resonant mode and ignoring the free body dynamics, one obtains
\[
G(s) := \frac{1}{6.6667 \times 10^{-8}}\TwoTwo{\frac{3.0907}{s^2 + 3.4^2} + 0.3}{\frac{3.557 \times 10^{-4}}{s^2 + 3.4^2}}{\frac{3.557 \times
    10^{-4}}{s^2 + 3.4^2}}{\frac{2.35}{s^2 + 3.4^2} + 0.3}.
\]
Note that $G$ is negative imaginary since $j[G(j\omega) - G(j\omega)^*] = 0$ for all $\omega \in (0, \infty) \setminus 3.4$. This follows from the
fact that $G(j\omega)$ is real and symmetric for all $\omega > 0$ such that $j\omega$ is not a pole of $G$. Furthermore, the residue matrix
\begin{align*}
  & \lim_{s \rightarrow j3.4} (s - j3.4) jG(s) \\
  = \, & \frac{1}{6.6667 \times 10^{-8}}\TwoTwo{\frac{3.0907}{6.8} + 0.3}{\frac{3.557 \times 10^{-4}}{6.8}}{\frac{3.557 \times 10^{-4}}{6.8}}{\frac{2.35}{6.8} + 0.3} > 0.
\end{align*}

To stabilise the plant $G \in \bs{N}$, an integral resonant controller (IRC)~\citep{PetLan10} is employed. An IRC is a first-order controller taking
the form
\[
\bar{G}(s) = (sI + \Gamma\Phi)^{-1}\Gamma - \Delta,
\]
which is strictly negative imaginary if $\Gamma > 0$, $\Phi > 0$ and $\Delta$ is symmetric~\citep[Thm. 8]{PetLan10}. Let
\begin{align*}
  \Gamma & := \TwoTwo{35}{15}{15}{20} \quad \Phi := \TwoTwo{745}{521}{521}{1.021} \\
  \Delta & := \TwoTwo{2.0871}{-1.0650}{-1.0650}{1.5229}.
\end{align*}
Note that since $\bar{G}(\infty) < 0$ and $G(\infty)\bar{G}(\infty) \neq 0$, \citep[Thm. 1]{XPL10} cannot be applied here to analyse the stability of
the feedback interconnection $[G, \bar{G}]$. However, it can be easily verified that $\bar{\sigma}(G(0)\bar{G}(0)) \approx 0$ and
$\bar{\sigma}(G(\infty)\bar{G}(\infty)) = 0.8720 < 1$, whereby Corollary~\ref{cor: img} holds and $[G, \bar{G}]$ is stable.

Suppose the feedthrough term of $\bar{G}$ is now 
\[
\Delta := \TwoTwo{10}{0}{0}{10}. 
\]
It follows that $\bar{\sigma}(G(0)\bar{G}(0)) = 3.3574 \times 10^7$ and $\bar{\sigma}(G(\infty)\bar{G}(\infty)) = 3$. As such, the conditions of
Corollary~\ref{cor: img} fail to hold and hence it cannot be used to conclude stability of $[G, \bar{G}]$. However, by defining
\[
\Pi_0 = \Pi_\infty := \TwoTwo{0}{I}{I}{0},
\]
it is straightforward to verify that the conditions in Theorem~\ref{thm: FB_ori} hold. As such, $[G, \bar{G}]$ is stable. Note that the multipliers
$\Pi_0$ and $\Pi_\infty$ employed in this example correspond to an IQC characterising passivity~\citep{MegRan97}. Intuitively, the stability of
$[G, \bar{G}]$ is established in this example by exploiting the fact that $\bar{G}$ and $G$ exhibit negative imaginary property at positive
frequencies that are bounded away from zero and infinity, and the positive real property when the frequencies are sufficiently small or sufficiently
large.

\subsection{Irrational transfer functions}

Consider $G = \frac{0.2}{s^2 + 1}$ and $\bar{G} = \frac{e^{-T s} + 3}{s+1}$. Observe that $\bar{G}$ contains an exponential term, which corresponds
to a time-delay operation in the time domain. It is straightforward to verify that $G \in \bs{N}$ and $\bar{G} \in \bs{N}_s$ for all time delays
$T \geq 0$. Furthermore, $\displaystyle\lim_{s \rightarrow j} (s - j) jG(s) = 0.1 > 0$, $G(0)\bar{G}(0) = 0.8 < 1$ and $G(\infty)\bar{G}(\infty) = 0$
for all $T \geq 0$. Therefore, by application of Corollary~\ref{cor: img}, it follows that $[\tau G, \bar{G}]$ is stable for all $\tau \in [0, 1]$ and
$T \geq 0$. Note that for this example, since $\bar{G}$ is an irrational transfer function due to the presence of the time-delay term, results
in~\citep{LanPet08, XPL10} are not applicable for concluding feedback stability.

Conversely, since it is known that $[\tau G, \bar{G}]$ is stable for all $\tau \in [0, 1]$ and $T \geq 0$, it follows from Theorem~\ref{thm: nec_IQC}
that there exist symmetric $\Pi_0$ and $\Pi_\infty$ such that the quadratic matrix inequalities therein hold. They can be taken, for example, to be
\[
\Pi_0 := \TwoTwo{0.8}{0}{0}{-1} \quad \text{and} \quad \Pi_\infty := \TwoTwo{0}{0}{0}{-1}.
\]

\section{Conclusions} \label{sec: con}

This paper establishes necessary and sufficient conditions for robust stability of feedback interconnections of negative imaginary
distributed-parameter systems using an integral quadratic constraint (IQC) approach. In contrast with the existing methods in the literature, the
results were obtained without exploiting explicit state-space or transfer-function representations. Of future interest are generalisations to
accommodate free body dynamics corresponding to poles at the origin~\citep{MKPL14}. Nonlinear systems exhibiting counterclockwise input-output
dynamics~\citep{Ang06} may also be considered within the framework of IQCs as extensions of negative imaginary systems to nonlinear settings.

\end{document}